# Low resistance NiO/β-$Ga_2O_3$ heterojunction diodes grown via molecular beam epitaxy

Dagny Sacksteder[1, 2], Anna Sacchi[1], Renae Gannon[1], Michelle Smeaton[1], Megan E. Holtz[2], Andriy Zakutayev[1,2], M. Brooks Tellekamp[1, 2]

[1]*National Laboratory of the Rockies, 15013 Denver West Parkway Golden, CO 80401, USA*
[2]*Colorado School of Mines, 1500 Illinois St. Golden, CO 80401, USA*

**Abstract**

NiO is one of the most important p-type oxide contact materials used in many semiconductor technologies. However, current NiO growth methods can induce interfacial damage that diminishes device performance. Fine control of interfaces is especially important in implementing NiO heterojunction diodes and transistors based on ultra wide band gap (UWBG) semiconductors such as AlGaN and $Ga_2O_3$ used for power electronic applications. Here, we report on how molecular beam epitaxy can be used to achieve low-defect, lightly doped NiO contact layers for a β-$Ga_2O_3$ diodes. Although high-temperature growth does not measurably decrease the on-state resistance of the diode, increased growth rates up to 600 nm/hr lower on-state resistance in $p^{--}$ NiO / β-$Ga_2O_3$ heterojunction diodes without reducing film quality. At a NiO growth rate of 380 nm/hr, unoptimized diodes with 35 nm thick $p^{--}$ NiO layers demonstrate a device-average specific on-state resistance of 1.46 Ω-$cm^2$ and an ideality factor of 1.46. Individual devices grown at this condition show specific on-state resistance as low as 25 mΩ-$cm^2$ with a rectification ratio of $2.7x10^6$. Scanning transmission electron microscopy imaging reveals the (100) NiO/ (100) β-$Ga_2O_3$ interface is coherent and atomically abrupt. These results open a new avenue to optimizing the NiO interface to produce robust, competitive kV-class power electronic devices based on β-$Ga_2O_3$ and other UWBG semiconductors.

## 1. Introduction

NiO is an archetype p-type conductive oxide, an important family of functional materials as contacts in electronic and optoelectronic applications. While its initial discovery in 1993 [1] was soon outshone by better ternary p-type oxides [2], the simplicity of binary chemistry and cubic crystal structure of NiO has maintained its status as one of the most widely used p-type contact across multiple semiconductor technologies - from silicon [3] and III-Vs [4], to organic [5] and hybrid [6] materials. Among wide band gap semiconductors, NiO has been used as a p-type contact to GaN for nearly three decades in a form of oxidized Ni contacts [7,8], while more recently sputter deposition of NiO on GaN has been shown to improve heterojunction performance [9] and edge termination [10], resulting in demonstration of high-breakdown voltage diodes [11] and high-electron mobility transistors (HEMTs) [12]. The remaining challenge in the field is to gain understanding and develop process to apply NiO and related p-type oxides to semiconductors without the damage caused by NiO deposition process.

β-$Ga_2O_3$ is a promising ultrawide bandgap material for power electronics applications with potential for scalable production due to bulk melt growth, high-quality epilayers, efficient n-type doping, and high theoretical breakdown voltage compared to SiC and GaN [13-15]. However, its implementation is impeded by the lack of p-type doping in β-$Ga_2O_3$ which necessitates heterointerfaces to produce p-n diode devices. NiO is a leading p-type oxide candidate due to its wide band gap of 3.77 eV and favorable band alignment with β-$Ga_2O_3$ [16,17]. While stoichiometric NiO is insulating, it can be intrinsically doped with Ni vacancies to induce p-type conductivity [18-20]. NiO/β-$Ga_2O_3$ devices have achieved breakdown voltages > 8 kV, with reverse leakage currents < 1μA/$cm^2$, and specific on-state resistance values < 20 mΩ-$cm^2$ [21-23]. These results motivate further research to fully optimize and control the NiO/β-$Ga_2O_3$ interface, enabling competitive power electronic devices.

Despite these promising demonstrations, NiO devices are impacted by interphase reactions [24] and long-term deterioration in the presence of $H_2O$ [25]. NiO/β-$Ga_2O_3$ devices in literature are most often grown using reactive radio-frequency (RF) sputtering [25-27], an energetic process which can introduce interfacial contamination and extended defects. One recent study measured electrically active damage ~10 μm into the β-$Ga_2O_3$ drift layer following the NiO sputter process, seeding premature breakdown and degrading the turn-on characteristics [28]. Other available techniques such as pulsed laser deposition [24,29,30] induce less damage than sputtering but are limited to small grains and high dislocation densities. Ultimately, such defects cause existing devices to fall short of predicted limits for breakdown voltage and specific on-state resistance ($R_{on,sp}$) in β-$Ga_2O_3$. Realizing competitive β-$Ga_2O_3$ heterojunction diodes (HJD) will require new device synthesis approaches with finer control over the structure and composition of the p-n interface to eliminate external failure modes and approach the intrinsic limits of β-$Ga_2O_3$. Recently, improved device performance was demonstrated by adding an 8 nm electron beam evaporated layer of NiO between the β-$Ga_2O_3$ (011) and sputtered $NiO_x$ layers, achieving > 3 kV breakdown and a power figure of merit value > 2.3 GW/$cm^2$ [22]. Evaporation techniques do not introduce energetic particles which can damage the β-$Ga_2O_3$ surface during deposition of the initial layers, thus forming a protective layer of $p^{--}$ NiO which can be covered with conductive sputtered $p^{++}$ $NiO_x$ to achieve a graded p-i-n junction.

In this work, we explore how NiO grown via molecular beam epitaxy (MBE) can tune the conductivity of $p^{--}$ NiO/(100) β-$Ga_2O_3$ to improve interfacial quality and device performance. NiO is lattice matched to (100) β-$Ga_2O_3$ in a cube-on-pseudocube epitaxial relationship that prohibits the formation of twin domains [31], allowing us to achieve low-defect, coherent interfaces. We employ two strategies to induce intrinsic p-type conductivity in NiO: 1) increasing the equilibrium vacancy concentration by growing at high temperatures under excess oxygen according to the expression $X_v = \exp(-\Delta G_v/RT)$ and, 2) kinetically trapping vacancies with accelerated growth rates in oxygen-rich conditions. High growth rates (between 230-600 nm/hr) achieved an increase in NiO conductivity, decreasing the average $R_{on,sp}$ to ~1 Ω-$cm^2$, with individual devices demonstrating specific on-state resistances < 30 mΩ-$cm^2$. Higher growth temperatures led to increasingly less conductive and rougher films with much lower reliability in device performance. These results demonstrate new and scalable ability to design diodes for optimal performance without compromising interfacial quality, unlocking new routes towards competitive NiO/β-$Ga_2O_3$ HJD devices with graded p-type layers.

## 2. Experimental

Films were grown on cleaved β-Sn:$Ga_2O_3$ (100) substrates ($N_D$ ranged from 1 x $10^{16}$ – 5 x $10^{17}$ $cm^{-3}$). The substrates were cleaved along the (100) plane prior to loading. To improve thermal absorption from the radiative heater, substrates were indium-bonded to silicon handles before loading. Samples were outgassed at a temperature 100 °C higher than the grown temperature for 10 minutes and then annealed in oxygen plasma at the growth temperature for 10 minutes prior to beginning growth. All films were grown in an oxygen partial pressure of approximately 2.5 x $10^{-5}$ mbar supplied by RF oxygen plasma powered at 250 W, and substrate temperatures between 500°C and 1000°C. Ni fluxes between 1.5 x $10^{14}$ atoms/$cm^2$-s and 1.2 x $10^{14}$ atoms/$cm^2$-s (100 nm/hr – 630 nm/hr) were supplied by an electron beam evaporator, enabling a molten source and higher growth rates than are possible with Ni in an effusion cell. Samples were characterized *in-situ* via reflection high energy electron diffraction (RHEED). The MBE growth conditions used in this study are summarized in Table S1 of the supplementary information. All samples are between 25 nm and 44 nm thick, allowing for comparison across the temperature and flux series.

Samples were evaluated for their phase purity and crystalline quality using high-resolution X-ray diffraction (XRD) with a monochromated Cu-Kα X-ray source. Thickness was determined using x-ray reflectivity (XRR) measurements (SI Figure 1). Atomic force microscopy (AFM) was used to determine surface roughness and qualitatively identify growth modes. We also conducted XRD rocking curve ω scans to quantify crystalline quality. However, twin domains in the substrates and mis-tilt induced by the imperfect cleaving process result in difficult-to-interpret rocking curves (SI Figure 2). Therefore, we relied on RHEED patterns and AFM height maps to distinguish film quality with respect to different growth conditions. Both of these techniques are surface sensitive; RHEED provides insight into surface roughness and crystallinity, while AFM indicates film surface morphology and a direct measure of roughness. Since device performance is highly dependent on avoiding interfacial features which can concentrate electric field and cause premature breakdown, we find the data available from RHEED and AFM to be good criteria for optimizing growth parameters and evaluating film quality.

An ultra-thin specimen for Scanning Transmission Electron Microscopy (STEM) was prepared using Xenon Plasma Focused Ion Beam (PFIB) and Scanning Electron Microscopy (SEM) on a Thermo Fisher Scientific Helios 5 PFIB CXe using standard PFIB lamella preparation techniques[32]. Before STEM specimen preparation, the NiO film surface was imaged with SEM (SI Figure 3). STEM imaging was performed on a Thermo Fisher Scientific Spectra 200 STEM operated at a 200 kV accelerating voltage with a 24.2 mrad convergence semi-angle. STEM electron energy loss spectroscopy (EELS) data were acquired using a Gatan Enfinium spectrometer. EELS maps were acquired in dual EELS mode, and the zero-loss peak was used to align each O-K edge spectrum in energy. The spectra were background subtracted using standard methods and subsequently binned to improve signal to noise ratio.

To fabricate NiO/$Ga_2O_3$ vertical heterojunctions from these films, a Ti/Au (5/100 nm) planar ohmic metal stack was deposited on the back side of the $Ga_2O_3$ wafer by a Temescal FC2000 electron beam evaporator, under high-vacuum conditions, covering the entire available surface area. Circular Ni/Au (30/100 nm) Schottky top contacts with diameters ranging from 200 μm to 500 μm were then deposited on top of the NiO epilayer, by e-beam evaporation through a physical shadow mask. Prior to both back- and front-contact deposition, the samples were exposed to 100 W UV Ozone for 10 minutes at room temperature to remove surface contaminants and promote a clean contact interface. J-V measurements were completed in a probe station under ambient conditions using a semiconductor parameter analyzer. Turn on voltage ($V_{on}$) was determined by interpolating the voltage at which the current density equaled 0.1 A/cm$^2$. To eliminate data variability from inconsistent contact quality, only devices with an ideality factor between 1.3 and 2.5 and a $V_{on}$ between 1.3 V and 3 V were considered in subsequent analysis. The low cutoff may exclude the best data, however it also prevents inclusion of accidentally shorted devices that behave like junction barrier Schottky (JBS) diodes. Specific on-state resistance was extracted from J-V data by determining the slope of the J-V curve after saturating the series resistance. Although the diodes in this study turn on and rectify well, the high resistivity of the NiO films made it difficult to extract C-V measurements.

## 3. Results and Discussion

### *3.1 Growth temperature alters growth mode and conductivity*

To explore whether growth temperature could induce meaningfully higher vacancy concentrations in NiO, we grew a temperature series between 500°C – 1000°C (at constant flux of 100 nm/hr and constant thickness of approximately 30 nm). The XRD 2θ-ω scans used to evaluate film quality and orientation across the temperature and flux series are shown in Figure 1.a. We note extraneous low-intensity peaks in some of the 2θ-ω scans; these are attributed to the $(4\ 0\ \bar{1})$ $Ga_2O_3$ peak at 31.66° and the forbidden (5 0 0) peak at 37.78° visible due to defects in crystallinity and the offcut of the substrate. The additional peaks at 30.60° and 32.94° in the samples grown at 600°C and 650°C are due to residual metallic indium from the substrate mounting. In all films, we observe a wide temperature window for phase-pure, epitaxially oriented (100) NiO, consistent with previous MBE studies on GaN and MgO [18,19]. All samples possess Laue oscillations except for the sample grown at 1000°C, indicating there is an upper bound in growth temperature for good crystalline quality NiO (100) films on β-$Ga_2O_3$.

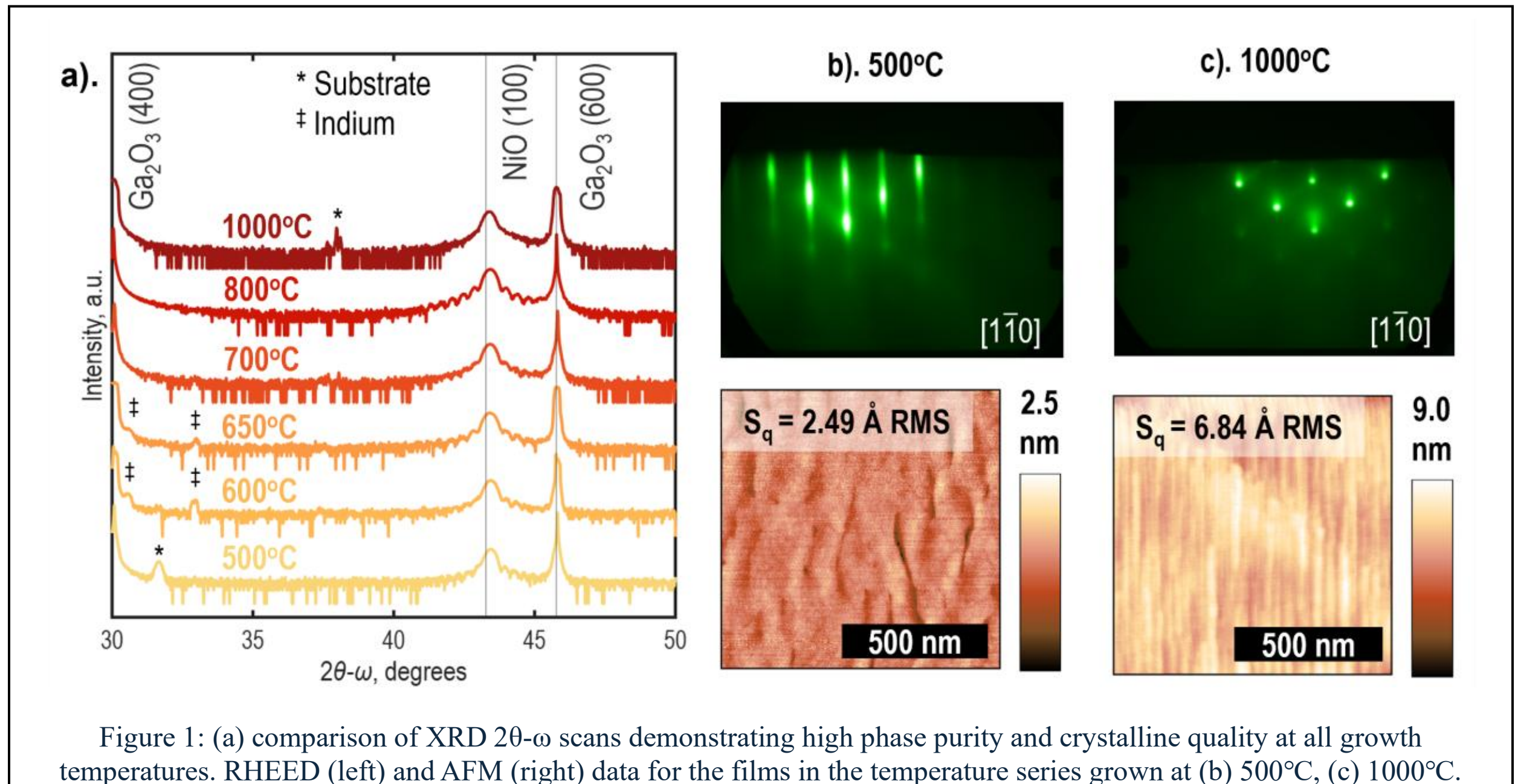

Figure 1: (a) comparison of XRD 2θ-ω scans demonstrating high phase purity and crystalline quality at all growth temperatures. RHEED (left) and AFM (right) data for the films in the temperature series grown at (b) 500°C, (c) 1000°C.

The RHEED and AFM data summarizing surface quality across the temperature series is presented in Figure 1.b-c. The RHEED pattern in Figure 1.b is the sharpest and most streaky, suggesting high crystalline quality and lower roughness at 500°C compared to the RHEED patterns of the films grown at higher temperatures. The RHEED pattern of the sample grown at 600°C (SI Figure 4) is spottier, and the samples grown at 650°C, 700°C, 800°C, and 1000°C (SI Figure 5, respectively) have significantly spottier and more diffuse RHEED patterns. The RHEED pattern of the highest temperature sample (Figure 1.c) has chevron patterns indicating crystal faceting along the surface. The spots in the higher temperature samples are indicative of rougher surfaces, while the more diffuse pattern originates from out-of-phase diffraction that indicates reduced crystallinity. This trend is supported by the AFM data. The sample grown at 500°C has the lowest roughness compared to the samples grown at higher temperatures. It also has a noticeably different surface morphology compared to the samples grown at higher temperatures; whereas the other films have indications of islands (SI Figure 4.a & 4.d), pin holes (SI Figure 4.b-c), or faceting (Figure 1.c), the AFM data in Figure 1.b supports a flat, island-free growth mode. This condition is ideal for maintaining low roughness as the film thickness increases over the course of a deposition, so we selected 500°C as the growth temperature for the flux series.

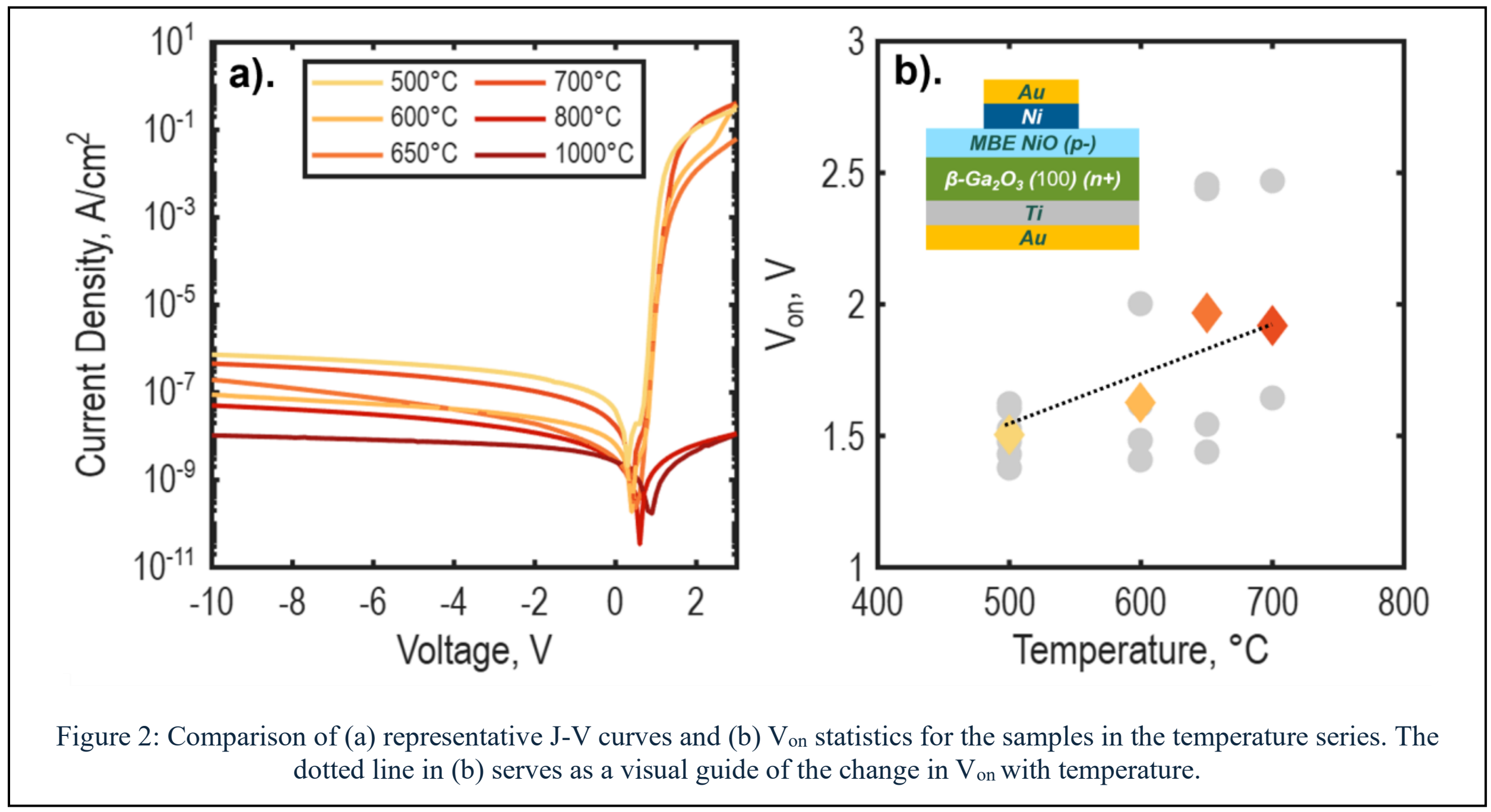

Figure 2: Comparison of (a) representative J-V curves and (b) $V_{on}$ statistics for the samples in the temperature series. The dotted line in (b) serves as a visual guide of the change in $V_{on}$ with temperature.

The findings from the J-V characterization of the samples in the temperature series are summarized in Figure 2. J-V curves from the best device at each growth condition are plotted in Figure 2.a. The samples grown below 700°C are rectifying, whereas the sample grown above 800°C were too insulating to turn on. A comparison of $V_{on}$ values for the temperature series is presented in Figure 2.b. Similar to the ideality factor and $R_{on,sp}$, we observe lack of a strong trend in $V_{on}$ with temperature. The samples grown at 500°C and 600°C have lower $V_{on}$ and less variability compared to the samples grown at 650°C and 700°C. SI Figure 5.a compares the ideality constants of all samples, revealing that although the lower temperature samples were rectifying, they have a wide range of ideality factors reflecting low control of device reliability. SI Figure 5.b reveals no relation between substrate temperature and the resistance of the NiO films. Overall, it appears growth temperature has an effect on film morphology but does not significantly change electrical performance in a repeatable and controllable way.

### *3.2 High growth rates increase NiO conductivity without compromising film structural quality*

Nickel vacancies have a lower energy of formation under oxygen-rich growth conditions [33]. However, due to the absorption-controlled growth of most high-quality binary oxide films with MBE, it is difficult to achieve a Ni-poor MBE film under oxygen rich conditions because the growth rate will be limited by the metal flux [34]. Instead, our approach to kinetic trapping of Ni vacancies in NiO leverages high Ni flux to achieve a high overall growth rate. By evaporating Ni with an electron beam, we were able to melt the source and achieve a much higher flux than a traditional effusion cell [18,19]. Higher growth rates at moderate temperatures would decrease the average diffusion length a Ni atom could travel prior to being pinned by subsequent layers, kinetically stabilizing a higher vacancy concentration in the film.

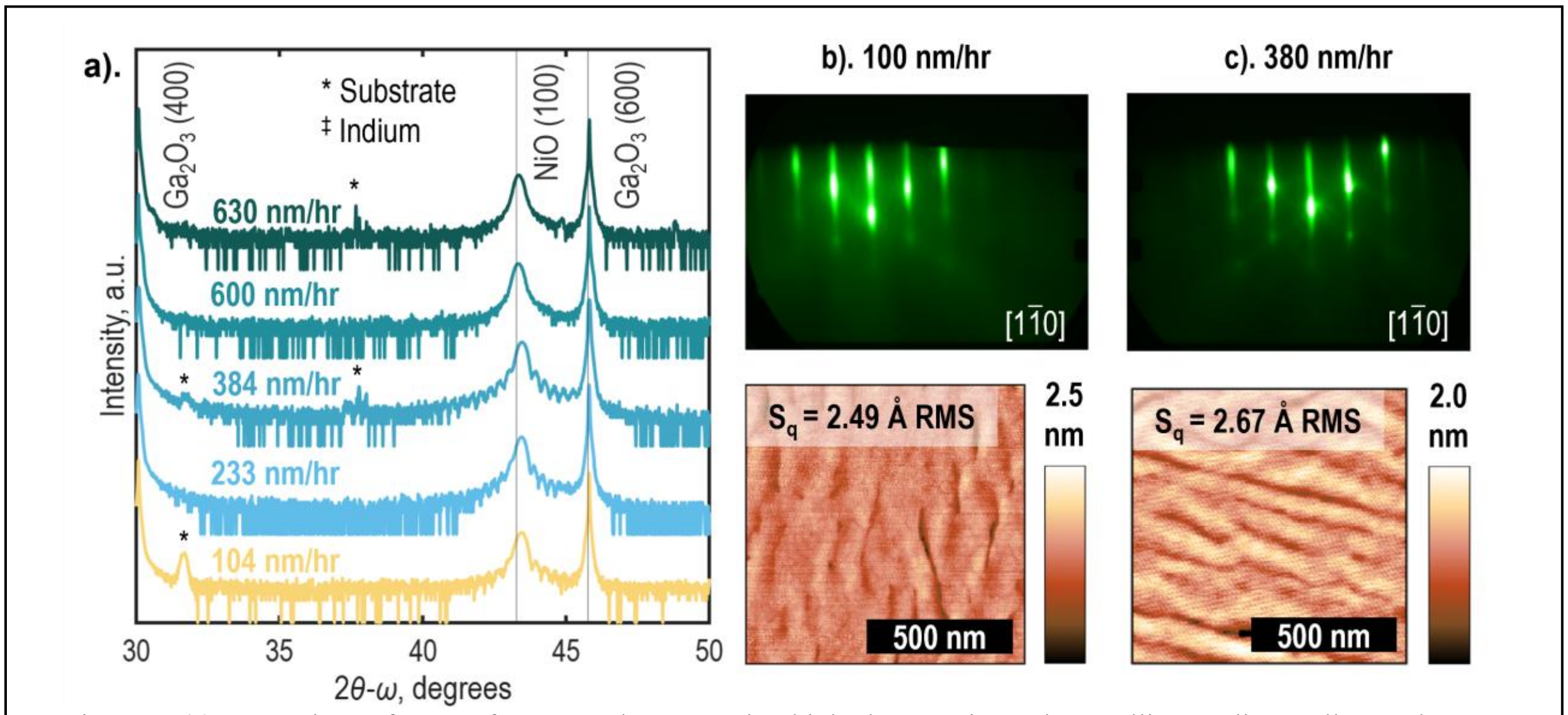

Figure 3: (a) comparison of XRD 2θ-ω scans demonstrating high phase purity and crystalline quality at all growth rates. Comparison of the RHEED and AFM data evaluating surface quality of the films grown at (b) 100 nm/hr and (c) 380 nm/hr.

The XRD, RHEED, and AFM data for the samples in the flux series grown at 500°C are summarized in Figure 3. In Figure 3.a, we observe films grown between 100 nm/hr and 380 nm/hr are phase pure (100) oriented NiO with Laue oscillations, indicating good epitaxial crystalline quality throughout a wide range of growth rates. The films grown at 600 nm/hr and 630 nm/hr were slightly thicker (40 nm according to XRR) compared to the rest in the series and did not have Laue oscillations. Some of these samples possess the same extraneous substrate peaks as described above in the temperature series (Figure 3.a). The peak near 45° in the 630 nm/hr sample could be strained Ni in the 1 1 1 out-of-plane orientation, suggesting there is an upper limit to high-flux growth at MBE background pressures before reaching a Ni-rich growth regime in which Ni metal begins to precipitate from the film. Comparisons of the RHEED and AFM data from the samples grown at 100 nm/hr and 380 nm/hr are presented in Figures 3.b-c, respectively (RHEED and AFM data from all other samples in the flux series are presented in SI Figure 6). All films appear to develop without island formation or faceting and maintain excellent surface roughness as growth rate increases. This suggests that at 500°C on cleaved (100) β-$Ga_2O_3$ surfaces, the Ni atomic surface diffusivity supports smooth growth up to a growth rate of at least 600 nm/hr without sacrificing film quality.

Next, we compared J-V characteristics of the samples in the flux series. Figures 4.a presents the J-V curves of the best devices in the flux series. We observe better leakage current at lower growth rates (100 nm/hr, 230 nm/hr, and 380 nm/hr), consistent with conductivity and defect concentration scaling with growth rate. Figure 4.b compares the $R_{on,sp}$ of the samples in the flux series. We observe a five-fold decrease in average $R_{on,sp}$ as the growth rate is increased from 100 nm/hr to 600 nm/hr, where the mean $R_{on,sp}$ was 94 mΩ-cm$^2$. The sample grown at 600 nm/hr had fewer rectifying devices, however out of devices that passed the screening described in the experimental section, this sample had the least variability in $R_{on,sp}$. At the highest flux condition (630 nm/hr), the $R_{on,sp}$ increased higher than the low flux condition, indicating an upper limit to kinetic trapping of defects, possibly due to the precipitation of metallic Ni in a Ni-rich growth condition. This is supported by the XRD and AFM structural characterization in Figure 3.b-c and SI Figure 7, which demonstrated a tradeoff with film quality at this highest growth rate. The comparison of the ideality constant for all the samples in the flux series follows the same trend as $R_{on,sp}$ (SI Figure 7.a). The $V_{on}$ has an inverse trend with growth rate compared to ideality and $R_{on,sp}$ (SI Figure 7.b). It follows that if higher growth rates are increasing the Ni vacancy concentration, the Fermi level in the NiO will move towards the valence band, increasing the built-in voltage. Therefore, these inversely related trends support the theory that kinetic trapping of Ni vacancies can induce intrinsic p-type doping of NiO.

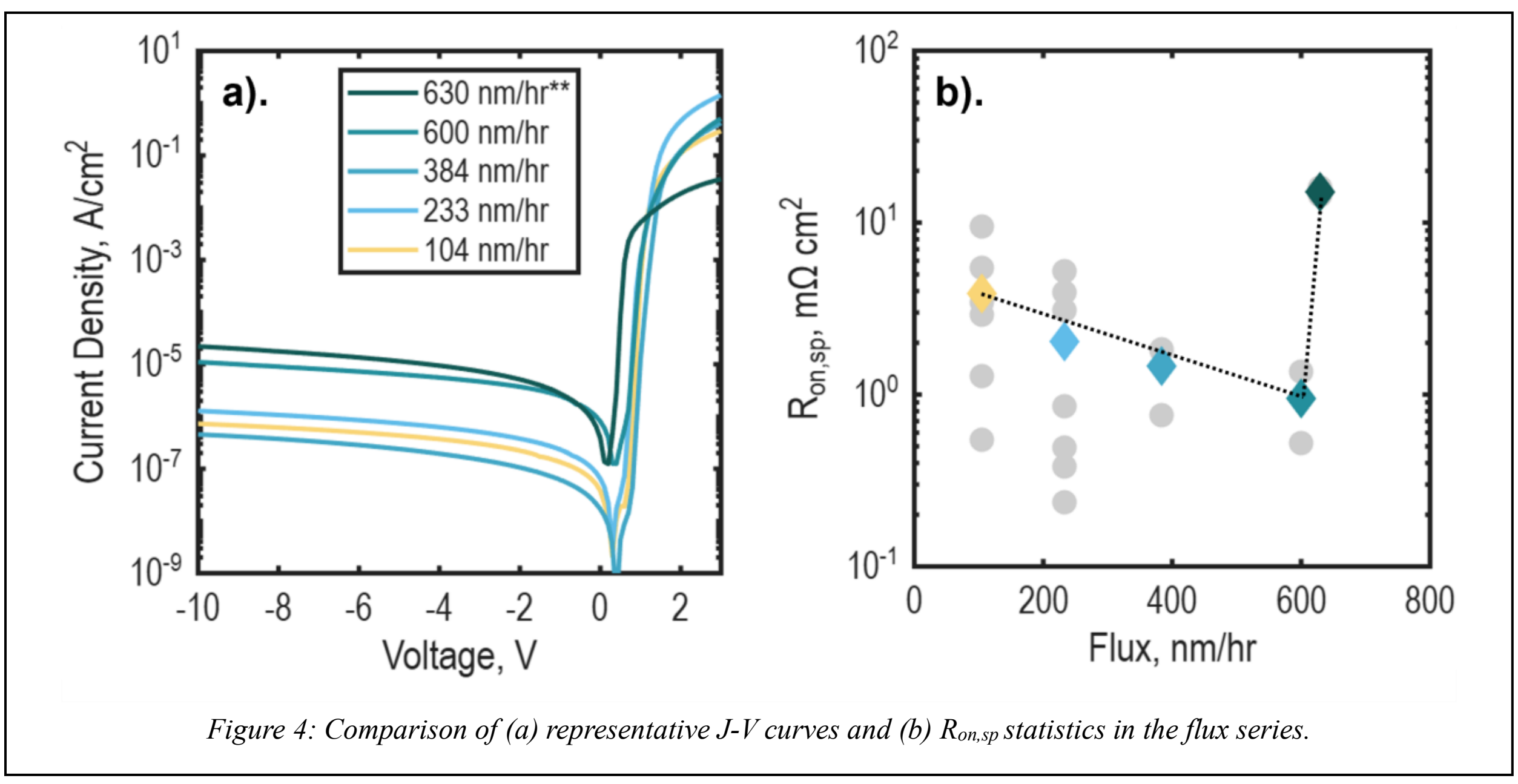

*Figure 4: Comparison of (a) representative J-V curves and (b) $R_{on,sp}$ statistics in the flux series.*

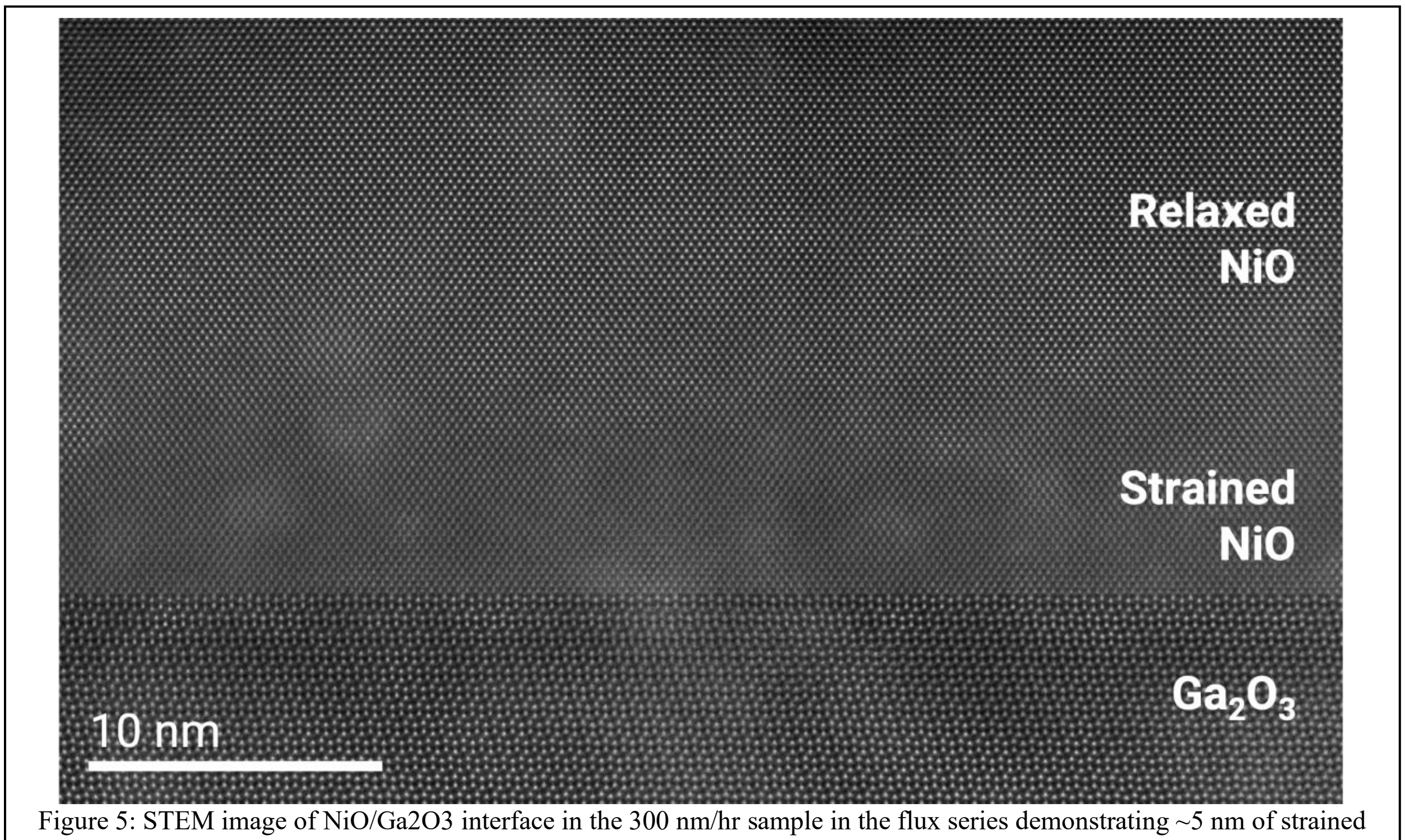

Figure 5: STEM image of NiO/Ga2O3 interface in the 300 nm/hr sample in the flux series demonstrating ~5 nm of strained NiO near the interface prior to relaxation.

Cross-sectional STEM imaging of the 380 nm/hr sample reveals homogeneous structure throughout the thickness of the film (Figure 5) with a sharp, coherent interface between the $Ga_2O_3$ and NiO. Some strain is observed in the first ~5 nm of NiO film, but minimal dislocations are present. After ~5 nm the film appears relaxed and nearly defect free. Furthermore, STEM EELS shows no variation in oxidation state through the thickness of the film, indicating even Ni vacancy incorporation (SI Figure 8). Based on these results, the 380 nm/hr condition appears to be the best compromise between $R_{on,sp}$, reverse leakage current, turn-on characteristics, overall ideality, and structural quality. Considering that these are shadow-mask deposited planar diodes without mesa etching, edge termination, or a graded NiO $p^{--}$-to-$p^{++}$ stack, we propose the measured device performance in these films will be significantly improved with deliberate processing. Ultimately, these results establish that higher growth rates can be used to kinetically trap nickel vacancies and achieve tunable p-type conductivity without compromising interfacial quality.

## Conclusion

Here, we have identified how MBE growth parameters alter the conductivity of NiO on cleaved (100) $\beta$-$Ga_2O_3$ substrates and tune device turn-on characteristics. We find that kinetic trapping of nickel vacancies can increase the p-type conductivity of NiO layers without compromising interfacial and film quality. At a growth rate of 380 nm/hr, the average device measured had a specific on-state resistance of a NiO/$\beta$-$Ga_2O_3$ p-n HJD was 1.46 $\Omega$ $cm^2$, ideality factor of 1.46. The best device grown at this condition achieved a specific on-state resistance of 25 m$\Omega$-$cm^2$ and a rectification ratio of $2.7x10^6$. Higher growth temperatures decrease film conductivity, structural quality, and overall device reliability due to the formation of pinholes, islands, and facets. Based on our characterization of surface quality and electrical performance, we identify a growth temperature of 500°C and a growth rate of ~400 nm/hr as ideal for growth of high quality, atomically flat and $p^{--}$ NiO films with promising electrical performance. Combined with state-of-the-art device design, this approach could enable significant reductions in processing-induced defects and failure modes in NiO HJD, pushing NiO/$\beta$-$Ga_2O_3$ devices towards the fundamental limits of the materials system.

## Acknowledgements

This work was authored in part by the National Laboratory for the Rockies (NLR) for the U.S. Department of Energy (DOE), operated under Contract No. DE-AC36-08GO28308. This work was primarily supported as part of A Center for Power Electronics Materials and Manufacturing Exploration (APEX), an Energy Frontier Research Center funded by the U.S. Department of Energy, Office of Science. Some of the AFM data included in this work was collected at the Colorado School of Mines Shared Instruments Facility Scanning Probe and Optical Microscopy: RRID:SCR_022048. D.S. acknowledges a graduate fellowship through the National Science Foundation Quantum Research Traineeship at the Colorado School of Mines. We also acknowledge assistance and useful discussion from: Henry Garland, Justin Rife, Stephen Glynn, and Dr. Yeageun Lee at NLR; Dr. Praveen Kumar at the Colorado School of Mines; Dr. Drew Haven at Excelitas; and Dr. Robert Lavelle at Penn State University.

**Low resistance NiO/β-$Ga_2O_3$ heterojunction diodes grown via molecular beam epitaxy**

Dagny Sacksteder[1, 2], Anna Sacchi[1], Renae Gannon[1], Michelle Smeaton[1], Megan E. Holtz[2], Andriy Zakutayev[1,2], M. Brooks Tellekamp[1, 2]

[1]*National Laboratory of the Rockies, 15013 Denver West Parkway Golden, CO 80401, USA*
[2]*Colorado School of Mines, 1500 Illinois St. Golden, CO 80401, USA*

## Supplementary Information

**Supplementary Table 1:** Summary of growth conditions for all samples included in study.

| Substrate doping | Substrate Temperature (°C) | Growth Rate (nm/hr) | Thickness (nm) | Notes |
|---|---|---|---|---|
| Sn | 1000 | 170 | 44 | Faceting |
| Sn | 800 | 120 | 25 | Laue oscillations |
| Sn | 700 | 120 | 28 | Laue oscillations |
| Sn | 650 | 120 | 25 | Laue oscillations |
| Sn | 600 | 120 | 32 | Laue oscillations |
| Sn | 500 | 100 | 26 | Laue oscillations, Smooth growth mode |
| Sn | 500 | 230 | 35 | Smooth growth mode |
| Sn | 500 | 380 | 32 | Smooth growth mode |
| Sn | 500 | 600 | 40 | Smooth growth mode |
| Sn | 500 | 630 | 38 | Ni precipitates |

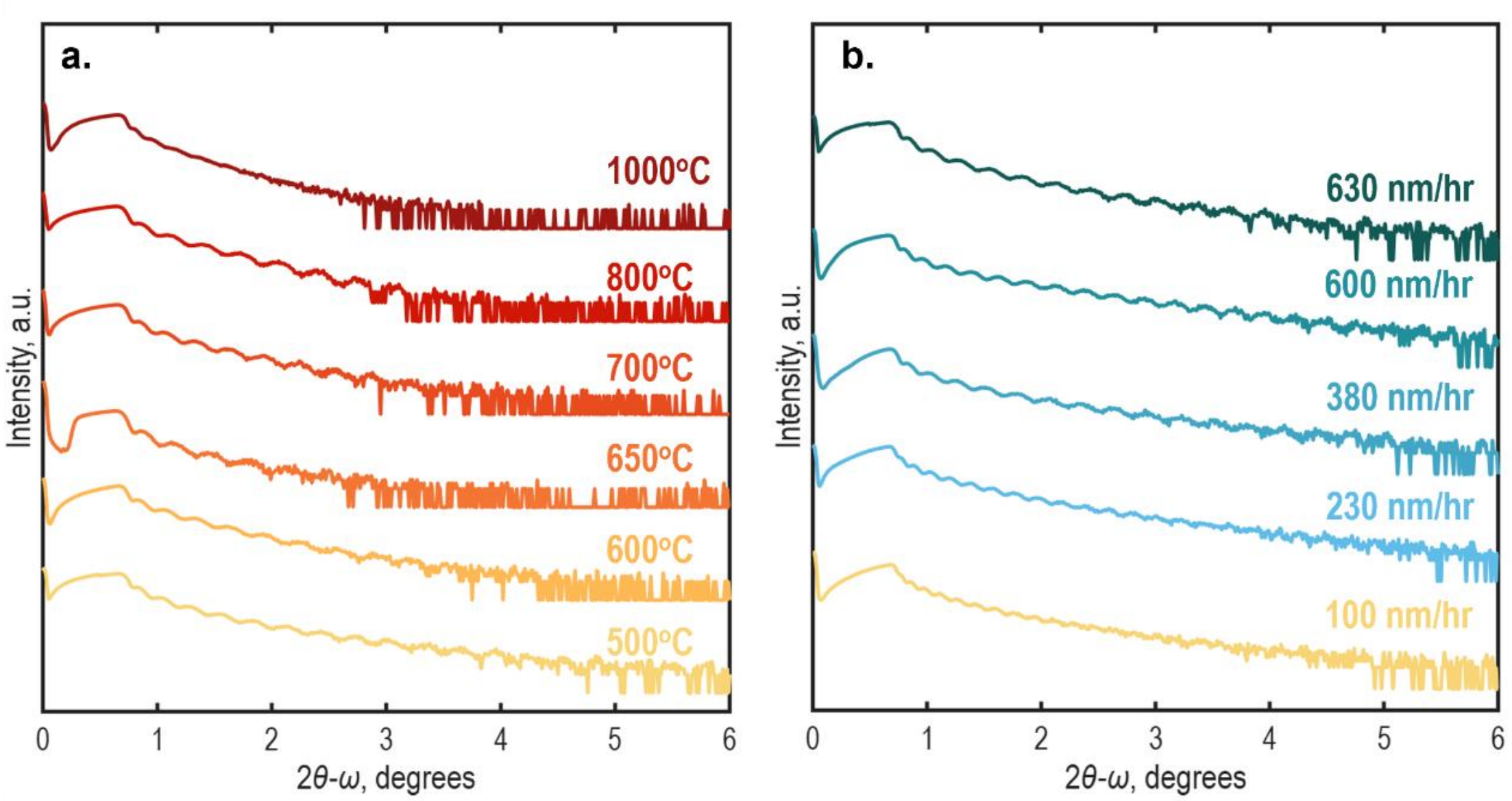

**Supplementary Figure 6:** X-ray reflectivity measurements used to determine film thicknesses listed in Supplementary Table 1.

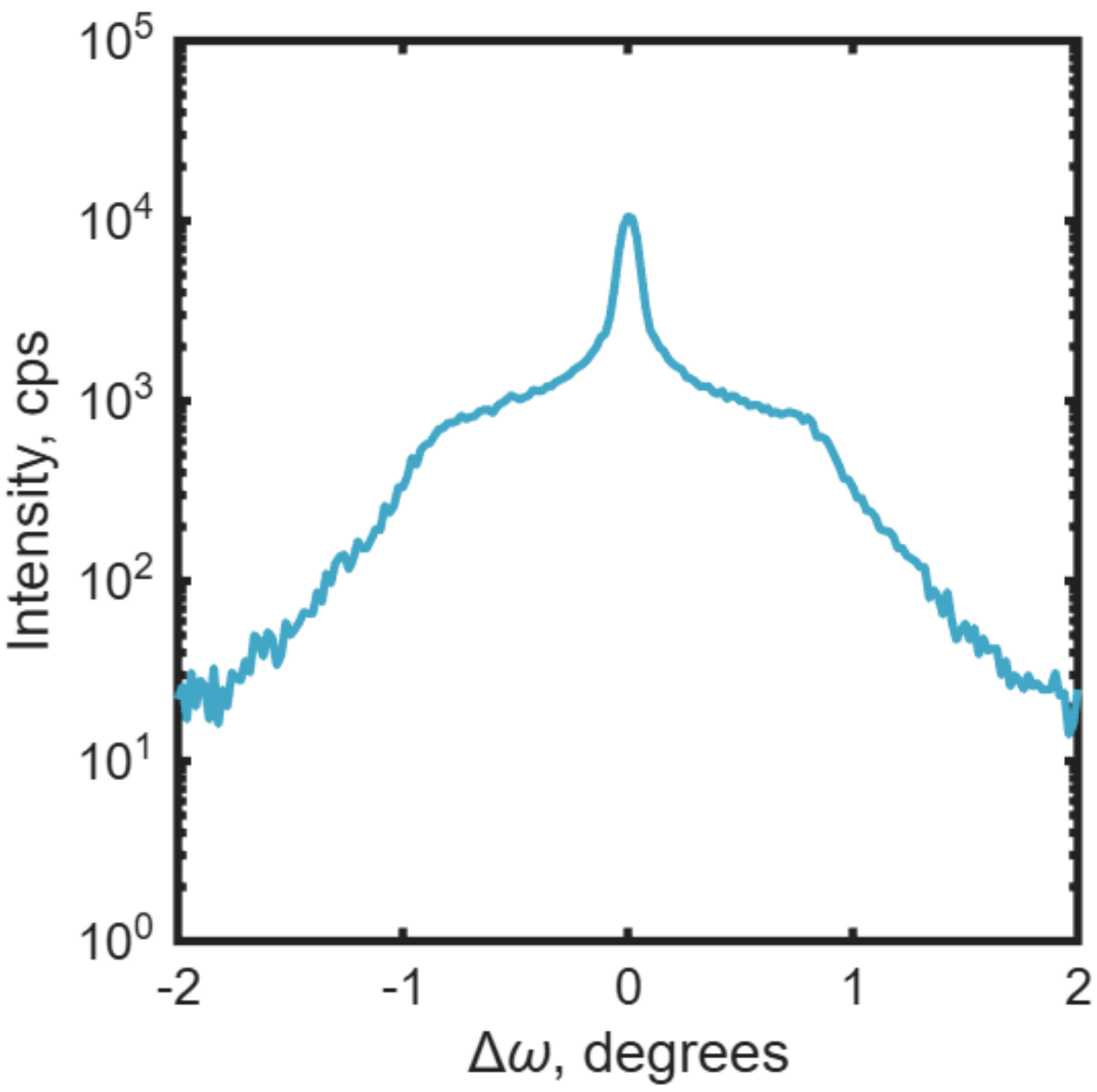

**Supplementary Figure 7:** Rocking curve of the sample grown at 380 nm/hr (representative of all samples used in this study). Varying substrate quality and flaking from the cleaving process led to difficult-to-interpret rocking curves from the film peaks.

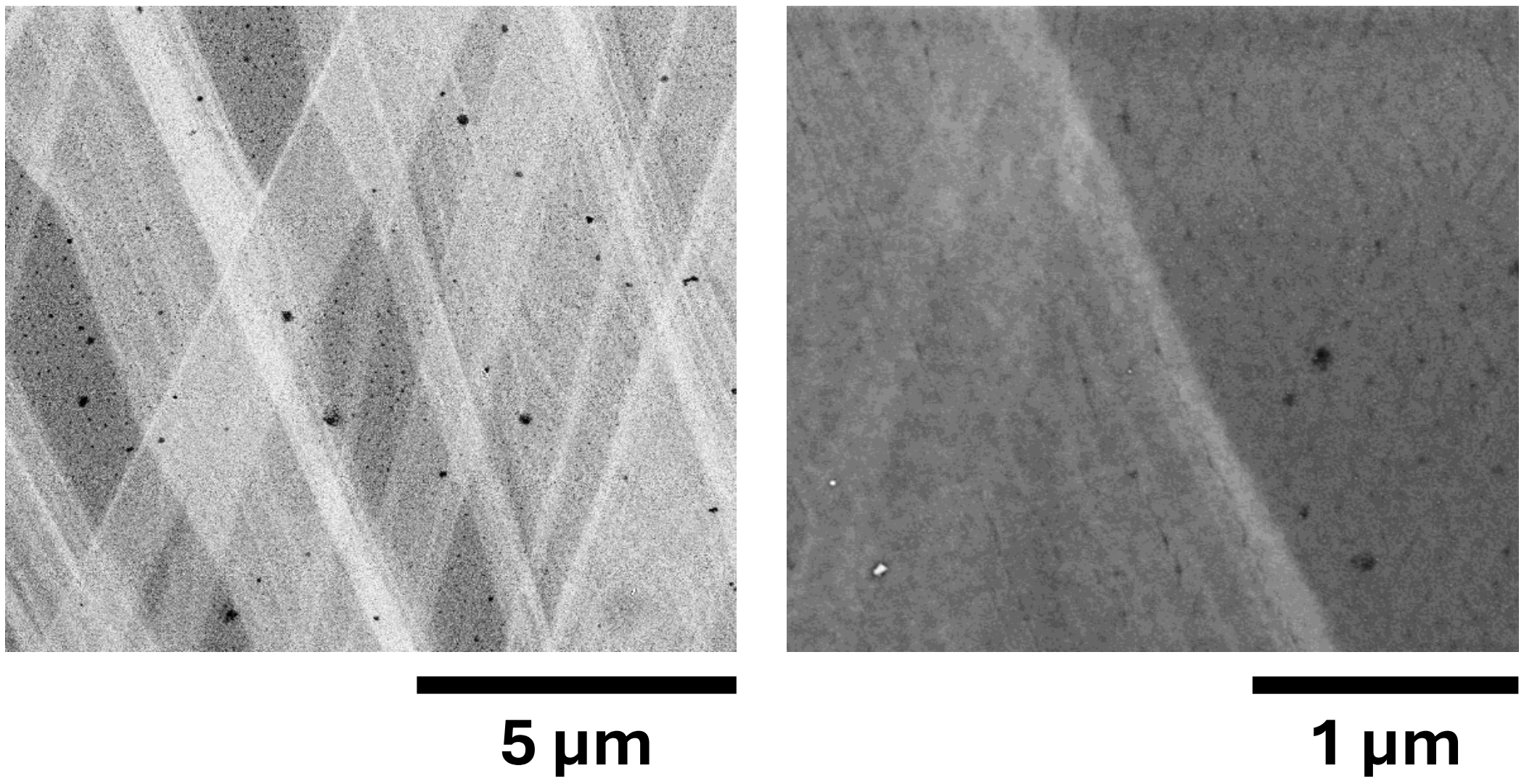

**Supplementary Figure 3.** SEM images of the sample grown at 380 nm/hr prior to FIB lamella preparation. Diagonal streaks are large terraces on the cleaved substrate surface.

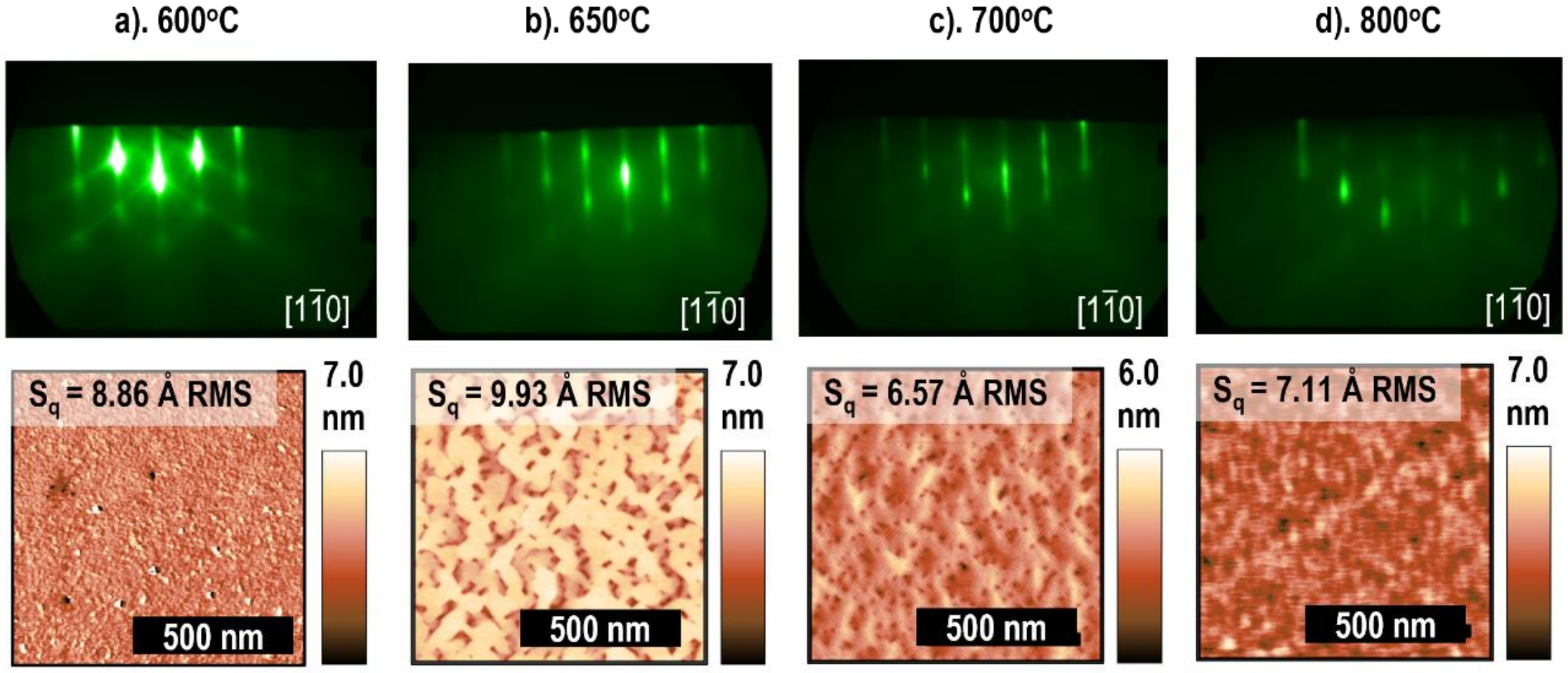

**Supplementary Figure 4.** RHEED and AFM of the samples from the temperature series grown at (a) 600°C, (b) 650°C, (c) 700°C, and (d) 800°C.

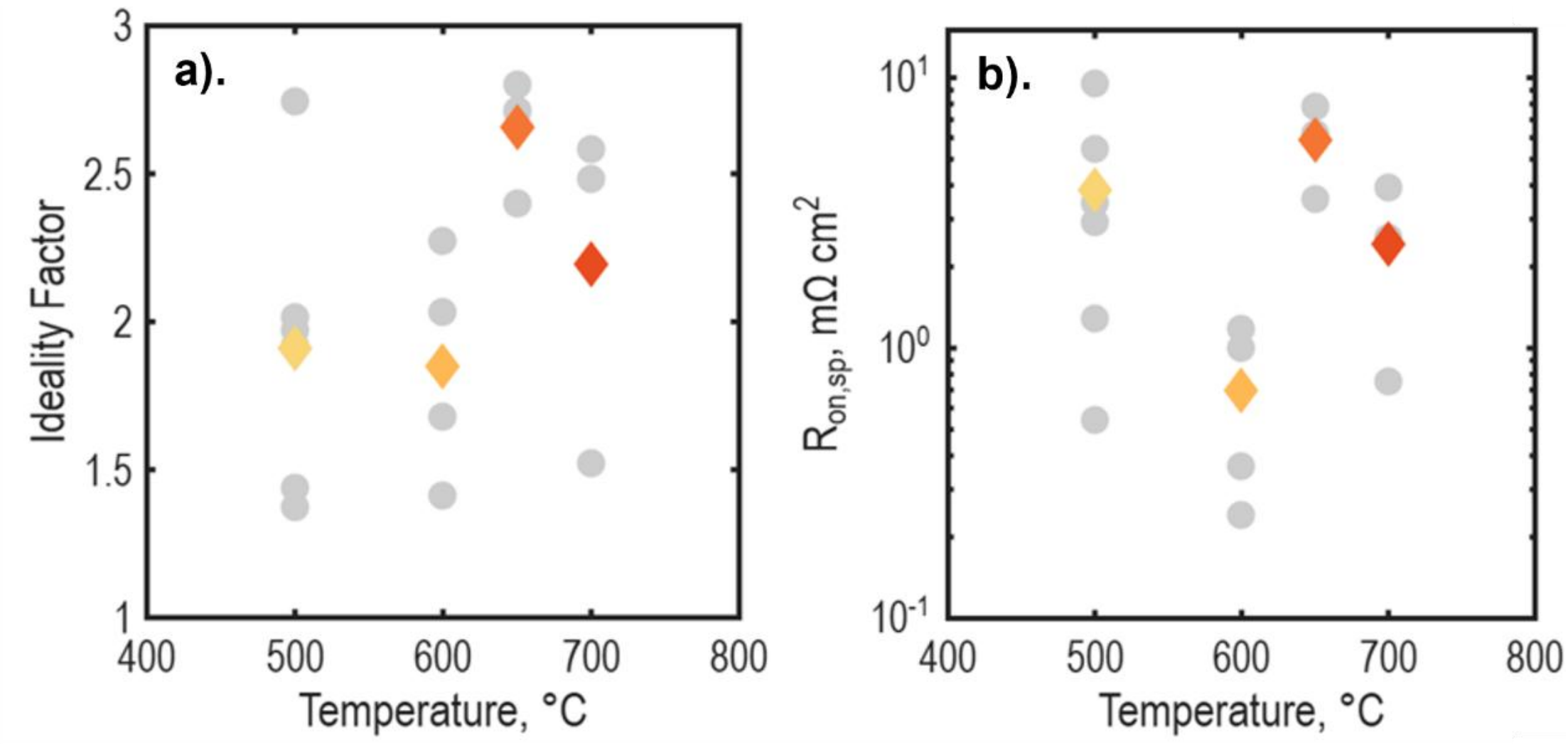

**Supplementary Figure 5.** (a) Comparison of the (a) ideality factors and (b) specific on-state resistance ($R_{on,sp}$) of the samples in the temperature series. Devices which passed the screening procedure described in the experimental section are plotted as gray circles. Average values for each device are plotted as colored diamonds.

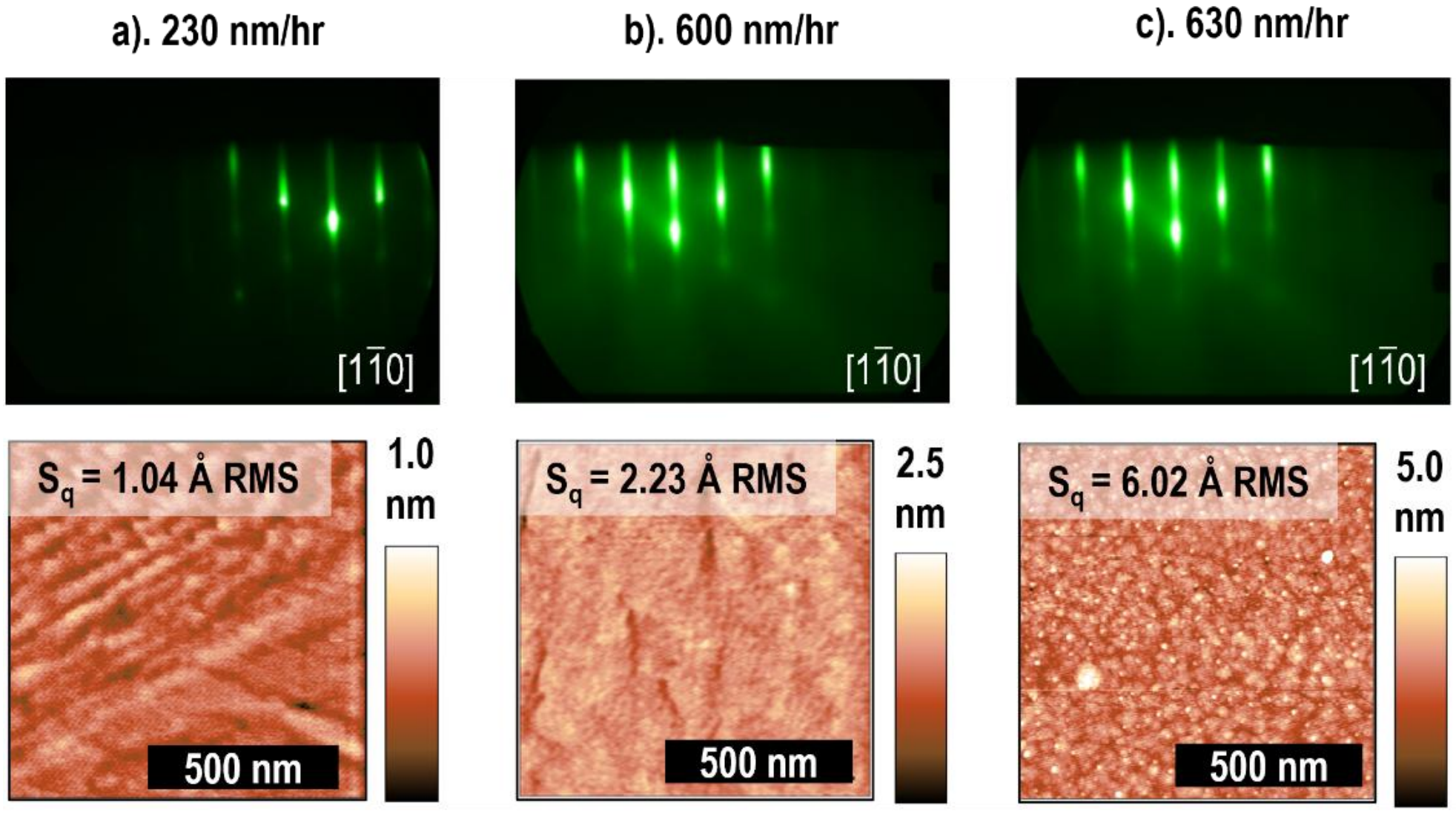

**Supplementary Figure 6.** RHEED and AFM of the samples from the flux series grown at (a) 230 nm/hr, (b) 600 nm/hr, and (c) 630 nm/hr.

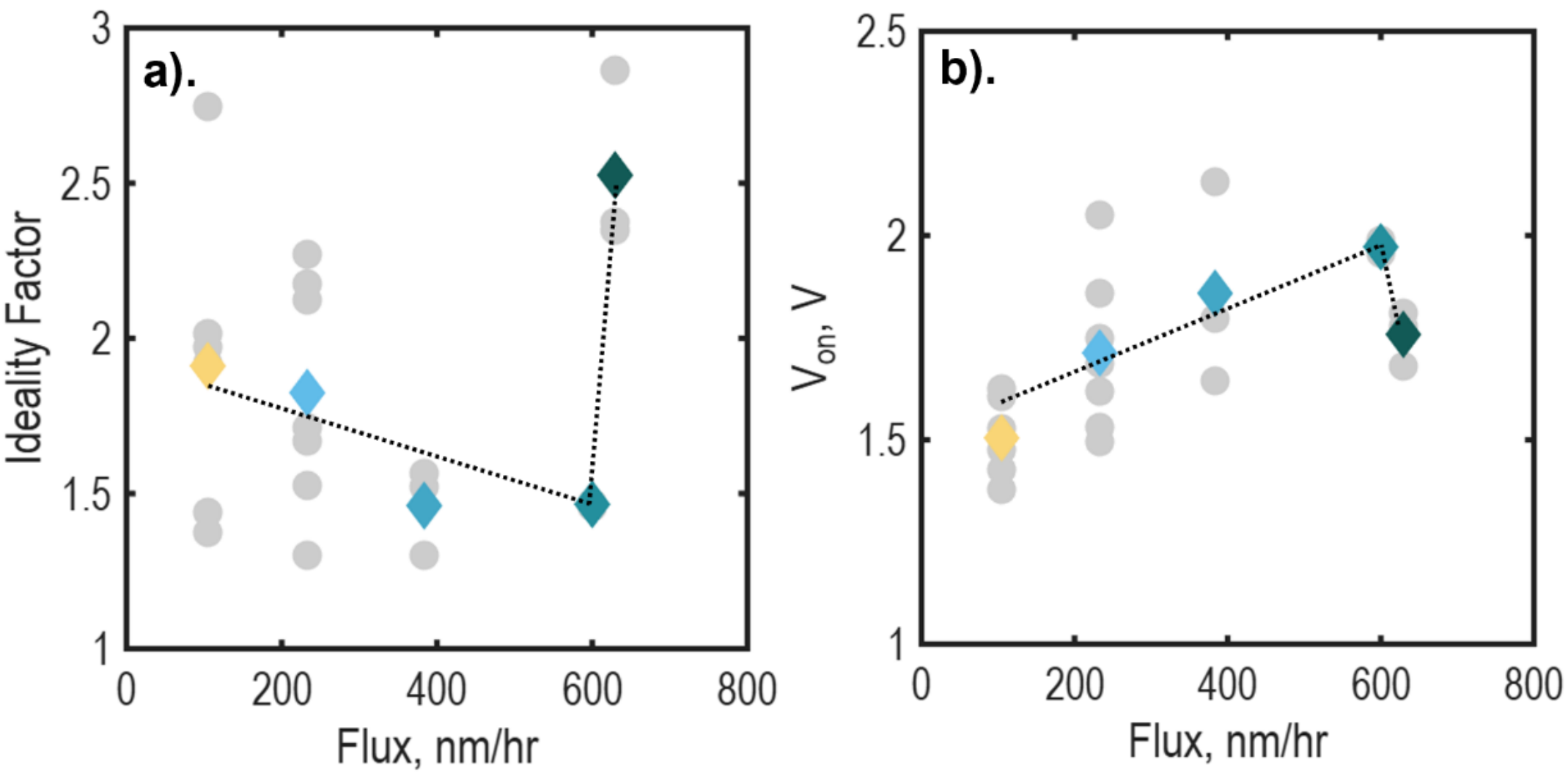

**Supplementary Figure 7.** (a) Comparison of the (a) ideality factors and (b) turn on voltage ($V_{on}$) of the samples in the flux series. Devices which passed the screening procedure described in the experimental section are plotted as gray circles. Average values for each device are plotted as colored diamonds.

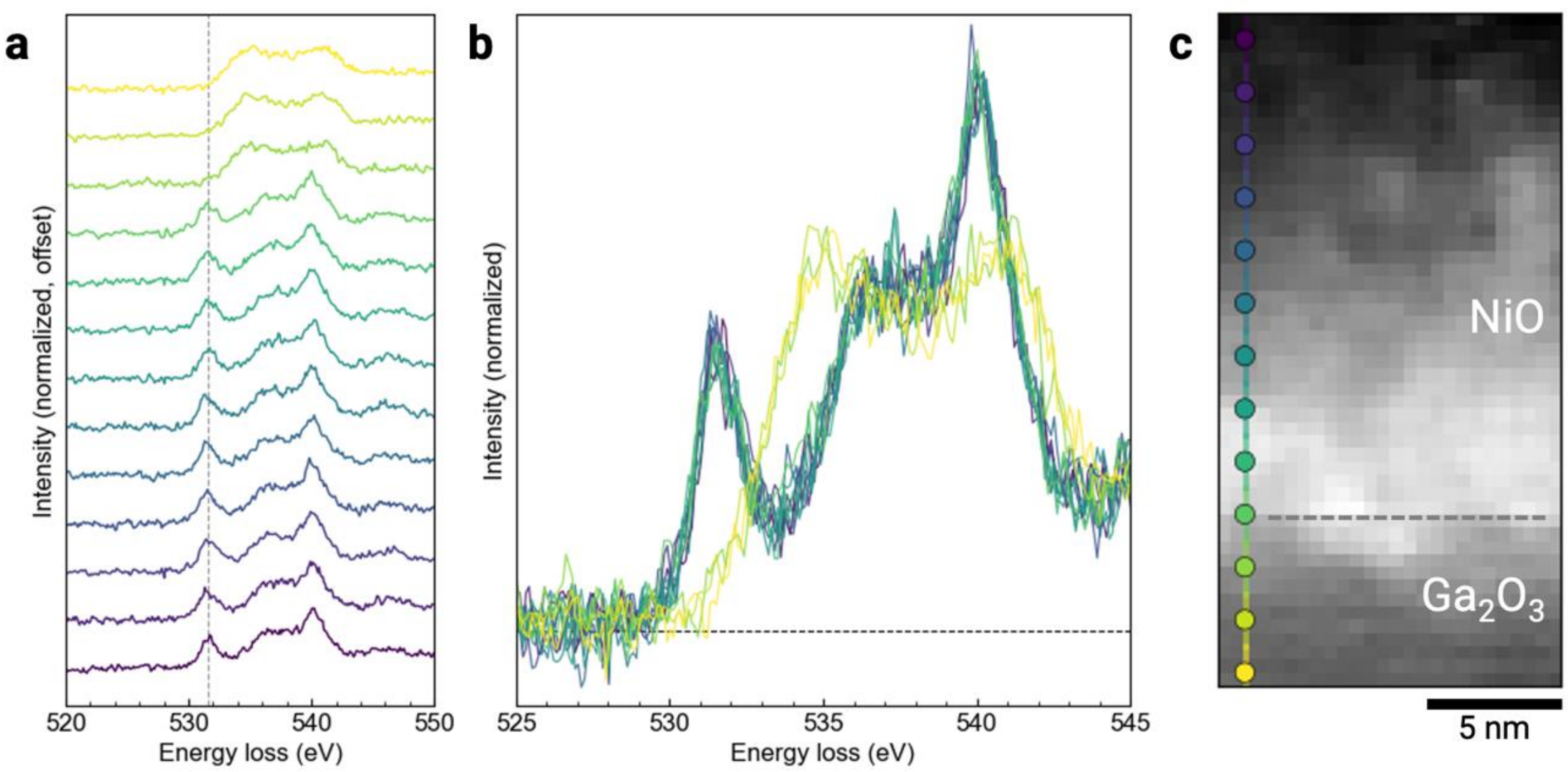

**Supplementary Figure 8.** STEM-EELS data demonstrating uniform composition throughout the NiO film grown at 380 nm/hr. (a) offset and (b) overlaid O-K edge spectra summed for the substrate/film layers indicated by colored circles in (c) the simultaneously acquired ADF image.